\documentclass[10 pt]{article}
\usepackage{graphicx}
\newcommand{\BEQ}{\begin{equation}}
\newcommand{\EEQ}{\end{equation}}
\newcommand{\BEA}{\begin{eqnarray}}
\newcommand{\EEA}{\end{eqnarray}}

\newcommand{\DQ}{{\prod_{a<b}^{1,n}{\rm d}Q_{ab} }}
\newcommand{\DLA}{{\prod_{a<b}^{1,n}{\rm d}\Lambda_{ab} }}

\newcommand{\dmza}{{\rm d}p(z_0)\,}
\newcommand{\dmzb}{{\rm d}p(z_1)\,}
\newcommand{\dmzc}{{\rm d}p(z_2)\,}

\newcommand{\mez}{\frac{1}{2}}

\newcommand{\s}{{\sigma}}
\newcommand{\tini}{{\mbox{\hspace*{-0.5 mm}\tiny $(i_1,i_2,i_3)$}}}
\newcommand{\nn}{{\nonumber}}

\begin{document}
    \title{The K-sat problem in a simple limit}
     \author{Luca Leuzzi$^{\star}$ and Giorgio Parisi$^{\star\star}$\\
 {\small $\star$ Instituut voor Theoretische Fysica, FOM,
Universiteit van Amsterdam}\\
{\small Valckenierstraat 65, 1018 XE Amsterdam (The Netherlands)}
{\small \tt leuzzi@wins.uva.nl}\\
 {\small  $\star\star$ Dipartimento di Fisica, INFM and INFN, Universit\`a di Roma}
   {\small {\em La Sapienza} }\\
{\small   P. A. Moro 2, 00185 Roma (Italy)}
{\small   \tt giorgio.parisi@roma1.infn.it }}
\maketitle
    \abstract{In this work we compute the thermodynamic properties of
 the 3-satisfiability 
    problem in the infinite connectivity limit. In this 
    limit the computation can be strongly simplified and the thermodynamic
 properties 
    can be obtained with a high accuracy. We find evidence for a continuous
replica symmetry breaking in the region of high number of  clauses,
$\alpha > \alpha_c$.}


 \section{Introduction}

 The statistical mechanics of the random $K$-satisfiability (K-SAT) problem
 has been the object of 
many studies in the last years \cite{MOZE1}\cite{nature}\cite{KISE}.
  The K-SAT was the first
problem to be shown to be Non-deterministic Polynomial (NP) complete
 \cite{COOK}.  This model
 is important because it provides a simple prototype for all the
 NP complete problems in 
complexity theory of computer science as well as in statistical mechanics
 of disordered and glassy 
systems, in computational biology and in other fields.

It is usually believed that the solutions of NP complete problems, or
 the certainty that they have 
no solutions, can only be found, in the worst case, by algorithms with 
a running time of computation 
that grows faster than polynomially (namely exponentially) with the number
 of variables $N$ of the 
system.

Generally speaking, in the statistical mechanics approach, for each
 given instance of the problem,
one introduces a Hamiltonian ${\cal{H}}(C)$,
constructed in such a way that the configuration $C^{*}$, which minimize
  ${\cal{H}}(C)$, is the solution 
of the problem if ${\cal{H}}(C^{*})=0$. On the contrary, 
if ${\cal{H}}(C)>0$ for any $C$, the problem 
does not have a solution.  In this framework one consider for each 
problem the partition function
\BEQ
Z(\beta)=\sum_{C}\exp(-\beta \cal{H}(C)) \ ,
\EEQ
where $\beta=T^{-1}$, $T$ being the temperature of the system.
 In the same way one introduces the 
usual thermodynamic quantities, e.g. the internal energy
\BEQ
{\cal{E}}(\beta)=-{ \partial \ln(Z(\beta))\over \partial \beta}
\EEQ

It can be argued that quantities like the internal energy density 
(i.e. $ E\equiv {\cal{E}}/N$) do not depend 
on $N$ in the infinite $N$ limit, so that a computation of 
their average over the different instances 
of the problem is sufficient for obtain interesting information 
in this limit.

When one studies the behavior of 
the K-SAT model at finite temperature, one finds a rich structure 
of phase transitions \cite{MOZE1}. In certain 
region of the parameter space, replica symmetry is broken
 (in other words there are many equilibrium 
states in the large volume limit \cite{mpv}). 
Explicit computation shows that it may be  possible to obtain  
a basic understanding of the connection between the 
${\mbox{SAT/UNSAT}}$ phase transition in random combinatorial 
structures \cite{MOZE1}\cite{nature} 
and the transition between a Replica Symmetric (RS) structure and a structure
 where the replica symmetry 
is broken in the frame of spin-glasses \cite{mpv}.
  Recent results \cite{BMW} suggest that  
the SAT/UNSAT transition 
seems to take place into the phase of broken replica symmetry.
 Yet the question stays open of how 
exactly the typical-case complexity theory of computer science 
and the Replica Symmetry Breaking 
(RSB) transition are related.

One of the main aim of the recent research on this model 
has been to understand better the structure of solutions, especially
at the borderline between the region of the phase diagram 
where the problem has solution and the region where no solution is possible.
Indeed this is the zone where the most unlikely (hardest) solutions are.

We will concentrate our attention on the case $K=3$ (3-SAT), that
is known to be the first and simplest NP complete instance
of K-SAT. The 2-SAT model is, in fact, already 
solvable in a time increasing polynomially (actually even linearly
\cite{linear}) with the number of 
variables. 

The basic boolean variables of the problem, $s(i)$, are defined on the sites 
$i$ with $i=1, ..., N$.
For 
technical reasons we prefer to use variables $\s(i)$ which take 
the values $\pm 1$. Our problem is then related to the original one  through
the variables transformation  
$s(i)=(1+\s(i))/2$. 

We consider an ensemble of randomly generated 3-SAT formulae. 
The Hamiltonian corresponding to a 
given formula is
\BEQ
{\cal{H}}=\sum_{i_1<i_2<i_3}r_{i_1,i_2,i_3}
\frac{1-\epsilon_{1}^{\tini}\s(i_1)}{2}
\frac{1-\epsilon_{2}^{\tini}\s(i_2)}{2}
\frac{1-\epsilon_{3}^{\tini}\s(i_3)}{2} \ .
\label{HAMILTONIAN}
\EEQ
For each instance of the problem we generate  $\alpha N$ clauses,
 where each clause 
is determined by randomly selecting three of the $N$ sites and
assigning to them a random $\pm 1$ variable. 
The terns of randomly chosen sites $\{i_1,i_2,i_3\}$ are given by
the variables $r_{i_1,i_2,i_3}$ that take the value $1$ with probability 
$p\equiv \alpha N^{-2}$ 
and the value $0$ with probability $1-p$. For finite $N$
there are 
approximately 
$\alpha N$ variables $r$ which are different from zero and they
 become exactly equal to $\alpha N$  in the limit $N\to \infty$.
Given a tern $\{i_1,i_2,i_3\}$, a set of three   variables 
$\epsilon$ are drawn, taking the 
value $+1$ or $-1$ with probability $1/2$.
 The function (\ref{HAMILTONIAN}) depends only on those variables
 $\epsilon^{(i_1,i_2,i_3)}$ such that
$r_{i_1,i_2,i_3}=1$ and  all the terms are non-negative:
${\cal{H}}$ just counts the number
of clauses that are not satisfied.
Obviously ${\cal{H}}=0$ if and only if all the clauses are satisfied.

As we have already explained we are going to consider  ${\cal{H}}$ as
 the Hamiltonian
of a disordered system,
in order to apply to the K-SAT model the statistical mechanics
techniques 
and to compute all the mathematical expressions in this framework.
We will introduce the fictive temperature  $T\equiv 1/\beta$
and at the end we will send $\beta\to\infty$ to
compute the ground state properties.

In the large $N$ limit we can  write the  equivalent Hamiltonian, in
 which the number of terms 
in the interaction is fixed and equal to $\alpha N$, as:
\BEQ
{\cal{H}}=\sum_{l=1,\alpha N}\frac{1-\epsilon_{1}^{(l)}
\s(i_{1}^{(l)})}{2}
\frac{1-\epsilon_{2}^{(l)}\s(i_{2}^{(l)})}{2}
\frac{1-\epsilon_{3}^{(l)}\s(i_{3}^{(l)})}{2} \label{seconda}
\EEQ
\noindent 
where the sites  $i_{t}^{(j)}$ ($t=1,2,3$) are randomly chosen for each one 
of the $\alpha N$  triplets.

For reasons that are discussed in \cite{MOZE1}\cite{nature}, one 
is interested
 to study the 
statistical properties of the system in the thermodynamic limit.  It is 
interesting to consider the zero temperature energy density $E_{0}(\alpha)$ 
(i.e. the average over 
the distribution of clauses of the number of clauses that are not satisfied
 by the formula corresponding to the Hamiltonian in equation 
(\ref{HAMILTONIAN})) as a function of the ratio $\alpha$ between the number
 of clauses and the 
number of variables.  We are eventually interested in the zero temperature
 entropy density 
$S_{0}(\alpha)$. The number 
of solutions satisfying 
the formula is asymptotically given by $\exp(N S_{0}(\alpha))$. 
 It has been conjectured 
\cite{MOZE1} that
\BEA
E_{0}(\alpha)=0, \ \ \  S_{0}(\alpha)>0,\ \  \mbox{for}\ \ \alpha<\alpha_{c} \ ,\\
S_{0}(\alpha)=0, \ \ \  E_{0}(\alpha)>0,\ \  \mbox{for}\ \ \alpha>\alpha_{c} \ ,\nonumber
\EEA
where the value of $\alpha_{c}$ is estimated to be around 4.2 \cite{KISE}.

Below $\alpha_c$ we have
 solutions (with probability going to 1 for $N$ going to infinity), 
while above $\alpha_c$ the problem
does not have solutions (i.e. it is UNSAT).  
At $\alpha \ll \alpha_c$ the problem is quite 
underconstrained and it is relatively easy to find an assignment 
of variables $\{\s_{i}\}$ 
satisfying the clauses.  For $\alpha \gg \alpha_c$, though 
in general still hard, to prove 
unsatisfiability is easier than in the hardest cases near $\alpha_c$. 
 Around the density of clauses 
$\alpha_c$ it is indeed very difficult either to find a satisfying
 assignment or to show 
unsatisfiability, i.e. it is most difficult to discriminate 
whether the problem admit any solution 
or no solution at all.  These are the cases where an exponential
 time may be needed.

Far from this critical value,
 anyway,  things simplify and more insight over the
structure of the phase space can be gained.

Indeed the exact evaluation of the free energy in the $\alpha$-$\beta$ plane
 is a rather complex 
computation. 
The aim of the present work is then to show that the computation strongly
 simplifies in 
the most overconstrained 
limit of $\alpha \to \infty$. Let us first introduce the  reduced 
inverse
temperature $\mu$ through the relation
\BEQ
\beta\equiv \frac{\mu}{\sqrt{\alpha}} \ . \label{def:beta_small}
\EEQ
 and let us define the rescaled energy density 
\BEQ
 e(\mu,\alpha) \equiv \frac{1}{\sqrt{\alpha}}\left(E(\beta,\alpha)-
	\frac{\alpha}{8}\right) \ ,\label{energy}
\EEQ
A similar definition can be written for 
the other 
thermodynamic functions. In particular for the free energy we have
the rescaled quantity
$ f(\mu,\alpha)=(F(\beta,\alpha)-\alpha/8)/\sqrt{\alpha}$. We shall 
also introduce the 
reduced temperature
$\tau=\mu^{-1}=T \alpha^{1/2}$.  
We will show below that the function $e(\mu,\alpha)$ has a 
limit when $\alpha$ goes to infinity. We can thus define 
\BEQ
e(\mu)=\lim_{\alpha \to \infty}e(\mu,\alpha).
\EEQ
In the following we will also compute the function $e(\mu)$ with high
 accuracy.
 In the  conclusions we will 
implicitly assume that the limit $\alpha \to \infty$ of full connectivity
and the zero temperature ($\mu \to \infty$) limit can be 
freely exchanged.

The interest for this computation is threefold:
\begin{itemize}
    \item The limit where $\alpha$ goes to infinity plays the same role of
 the infinite 
    connectivity model for spin glasses (finite connectivity/dilute 
models  correspond to finite $\alpha$) and 
most of our analytic understanding comes from the  study
 of the infinite     range models (Sherrington-Kirkpatrick like 
 models \cite{SK}), 
where the analytic computations are much simpler.
    \item
    Replica symmetry is broken in a region of the $\alpha$-$\beta$ plane,
 for $\alpha>\alpha_c\simeq 4.2$ \cite{KISE} 
 (or maybe for $\alpha > \alpha_s\simeq 3.9$ as recently
    derived in \cite{BMW} by means of a variational approach). 
 It is reasonable to assume that in 
    this whole region the way  in which replica symmetry is broken 
 is the same as in the limit $\alpha 
    \to \infty $.
    \item
    If we neglect the dependence of $e_{0}(\alpha) \equiv
 \lim_{\mu \to \infty}e(\mu,\alpha)$ on 
    $\alpha$ for $\alpha\ge\alpha_{c}$ (i.e. if we perform an
 asymptotic expansion in 
    $1/\sqrt{\alpha}$ and we consider only the leading order),
 we get the following estimate for 
    $\alpha_{c}$: 
    \BEA
    \alpha_{c}\approx(8 e_{0})^{2} \ \label{STIMA}.
    \EEA
Where the estimate $\alpha_{c}$ depends from the order of the 
 asymptotic expansion. In this work we will 
limit ourselves to the leading order in the $\mu^{-1}$ 
expansion. The precise value of $e_0$ will 
be given in the next section where we derive the thermodynamic observables 
using the replica tool.
\end{itemize}


\section{The Replica Formalism}
In the replica formalism one computes
\BEQ
Z^{(n)}\equiv\overline{Z[\{\Delta\}]^n} \ ,
\EEQ
where $\{\Delta\}$ denotes the random couplings,
 the bar is the average 
over  the distribution of the random couplings
 and the partition function is defined as:
\BEQ
Z[\{\Delta\}]=\sum_{\{\s_i\}} \exp(-\beta H[\{\s_i\},\{\Delta\}]) \ .
\EEQ
In our case $\Delta$ represents the ensemble of random
clauses, namely
the ensemble of terns $\{i_1,i_2,i_3\}$ with associated $\epsilon$'s.
The free energy density at finite $n$ is defined as
\BEQ
F^{(n)}=-\lim_{N\to\infty}\frac {\ln\left({\overline{Z^{(n)}}}\right)
}{\beta n N}\ .
\EEQ
\noindent where we firstly perform the  thermodynamic limit keeping 
$n$ fixed and only afterwards
 we send  $n\to 0$ by an analytic continuation procedure.
 
We are eventually interested in computing the limit $n \to 0$ of $F^{(n)}$,
 which is the 
value of the free energy density of the generic system in the infinite volume
 limit:
\BEQ
F=\lim_{n\to 0}F^{(n)}=-\lim_{N\to\infty}{\overline{\ln Z[\{\Delta\}]
}\over{\beta N}}\ .
\EEQ

Starting from (\ref{HAMILTONIAN}) and carrying out the average over 
the distribution of the $r$'s 
 we have 
\BEQ {\overline{Z^{(n)}}}=\overline{\sum_{\{\s_i^a\}}
\prod_{i_1,i_2,i_3}\left( 1-p+p \exp\left(-\beta\sum_{a=1,n}
\prod_{j=1,3}
\frac{1-\epsilon_{j}^{(i_1,i_2,i_3)}\s^{a}(i_j)}{2}
\right)\right)} \ .
\EEQ
\noindent where here the ${\overline{(...)}}$ is now the average only over the
$\epsilon$'s distribution. 
In the limit of large $N$ we can write the previous expression in terms
of the  effective Hamiltonian ${\cal{H}}_{eff}$ and the temperature
like parameter $\alpha$:
\BEQ
{\overline{Z^{(n)}}}=\sum_{\{\s_i^a\}} \exp\left(-\alpha {\cal{H}}_{eff}\right) \ ,
\EEQ
\noindent where
\BEQ
{\cal{H}}_{eff}\equiv\frac{1}{N^2}\sum_{i_1<i_2<i_3}
h_{eff}\left(\s(i_1),\s(i_2),\s(i_3)|\beta\right) \label{def:Heff}
\EEQ
\noindent and
\BEQ
h_{eff}\equiv 1 - {\overline{\exp\left(-\beta\sum_{a=1,n}\prod_{j=1,3}
\frac{1-\epsilon_{j}^{(i_1,i_2,i_3)}\s^{a}(i_j)}{2}\right)}}.
\EEQ
\noindent  For a given tern of sites $\{i_1,i_2,i_3\}$
the average is performed over the $2^{3}$ possible 
values of the three  variables $\epsilon$. 

Let us now consider the
 limit $\alpha \to
\infty$ at fixed $\mu$. The computation is long  but  straightforward.
In this limit the inverse temperature $\beta$, 
defined as in (\ref{def:beta_small}),
becomes a quantity of order $\alpha^{-1/2}$ so that we can 
freely expand the exponential in the previous expression (\ref{def:Heff}):
\BEQ
{\cal{H}}_{eff}=\frac{1}{N^2}\sum_{i_1<i_2<i_3}\left(
\frac{\mu}{\sqrt{\alpha}}{\overline{\sum_{a=1,n}\prod_{j=1,3}
\frac{1-\epsilon_{j}^{(i_1,i_2,i_3)}\s^{a}(i_j)}{2}}}
+\frac{\mu^2}{2\alpha}
{\overline{\left(\sum_{a=1,n}\prod_{j=1,3}
\frac{1-\epsilon_{j}^{(i_1,i_2,i_3)}\s^{a}(i_j)}{2}\right)^2}} +
 {\cal{O}}\left(\frac{\mu^{3}}{\alpha^{3/2}}\right)
						\right) \ ,
\EEQ
\noindent where
\BEA
&&{\overline{\sum_{a=1,n}\prod_{j=1,3}
\frac{1-\epsilon_{j}^{(i_1,i_2,i_3)}\s^{a}(i_j)}{2}}}=\frac{n}{8} \ ,\\
&&{\overline{\left(\sum_{a=1,n}\prod_{j=1,3}
\frac{1-\epsilon_{j}^{(i_1,i_2,i_3)}\s^{a}(i_j)}{2}\right)^2}}
=\frac{1}{8}\sum_{a,b}\prod_{j=1,3}\frac{1+\s^a_{i_j}\s^b_{i_j}}{2}=
\frac{n}{8}+\frac{1}{32}\sum_{a<b}\prod_{j=1,3}
\left(1+\s^a_{i_j}\s^b_{i_j}\right)
\EEA

Using standard manipulations, for large $N$ and $\alpha$, we easily get
\BEA
Z^{(n)}\hspace*{-3mm}&=&\hspace*{-3mm}
\exp{\left(-\frac{Nn\mu\sqrt{\alpha}}{8}+\frac{Nn\mu^2}{16}\right)}
\sum_{\{\s_i^a\}}\exp{\left(\frac{N\mu^2}{64}
\sum_{a<b}\left(1+\frac{1}{N}\sum_i\s_i^a\s_i^b\right)^3\right)}= 
\\
\hspace*{-3mm}&=&\hspace*{-3mm}\exp{\left(-\frac{Nn\mu\sqrt{\alpha}}{8}+
\frac{Nn\mu^2}{16}\right)}
\sum_{\{\s_i^a\}}\int \DQ \delta(\sum_i\s_i^a\s_i^b-N Q_{ab})
\exp{\left(\frac{N\mu^2}{64}\sum_{a<b}\left(1+Q_{ab}\right)^3\right)}=
\nonumber
\\
\hspace*{-3mm}&\equiv&\hspace*{-3mm} \int \DLA \DQ \exp(N n A[\{Q\},\{\Lambda\}]) 
\ ,\nonumber
\EEA

\noindent where
\BEQ
A[\{Q\},\{\Lambda\}] =\hspace*{-0.3mm}-\hspace*{-0.3mm}
\frac{\mu\sqrt{\alpha}}{8}\hspace*{-0.3mm}+\hspace*{-0.3mm}\frac{\mu^2}{16}
\hspace*{-0.3mm}+\hspace*{-0.3mm}
\frac{\mu^2}{64 n}\sum_{a<b}\left(1+Q_{ab}\right)^3
-\frac{1}{n}\hspace*{-0.3mm}\sum_{a<b}\Lambda^{ab}Q_{ab}\hspace*{-0.3mm}+
\frac{1}{n}\log \left(\sum_{\{\s_i^a\}}\exp \left(
\sum_{a<b}\Lambda^{ab}\s_a\s_b\right)\right) \ .
\EEQ	   

In the infinite $N$ limit we can use the saddle point method and we find
that the free  energy density is given by 
\BEQ
F(\alpha)= -\frac{ \sqrt{\alpha}}{\mu}
A[\{Q^{sp}\},\{\Lambda^{sp}\}] 
\EEQ
\noindent and the expression of the internal energy comes out to be:
\BEQ
E(\mu,\alpha)=\frac{\alpha}{8}-
\frac{\mu\sqrt{\alpha}}{8 } \left(1+\frac{1}{4 n}
\sum_{a<b}\left(1+Q_{ab}\right)^3\right) \ .
\label{ENERGY}
\EEQ

The elements
$\{Q^{sp}\}$ and $ \{\Lambda^{sp}\}$ satisfy the following consistency 
equations:
\BEQ
\Lambda^{sp}_{ab}=\frac{3\mu^2}{64}(1+Q^{sp}_{ab})^2\ , \ \ \ \ \
Q^{sp}_{ab}=\frac{\sum_{\{\s_a\}}\s_a\s_b\exp{\left\{\sum_{a<b}\s_a\s_b
\Lambda^{ab}\right\}}}
{\sum_{\{\s_a\}}\exp{\left\{\sum_{a<b}\Lambda^{ab}\s_a\s_b\right\}}} \ .
\EEQ

Our task is now to find the solution of these equations.

We notice that equation (\ref{ENERGY}) implies  that the exact definition of
 $e_0$ in (\ref{STIMA}) is 
\BEQ
e_0\equiv -\lim_{\mu \to \infty} \frac{\mu}{8}
\left[1+\frac{1}{4 n}\sum_{a<b}\left(1+Q_{ab}\right)^3\right]
\EEQ

\section{The Replica Symmetric Solution}
The  simplest  possibility (which is correct at high temperature) 
consists in assuming 
that the off-diagonal elements of the matrix $Q$ and $\Lambda$ are 
constant and they are 
equal to $q_{0}$ and $\lambda_{0}$ respectively. A simple computation shows 
that
\BEA
{A}(q_0,\lambda_0)&=&-\frac{\mu\sqrt(\alpha)}{8}+\frac{\mu^2}{16}-
\frac{\mu^2}{128}(1+q_0)^3-\frac{\lambda_0 (1-q_0)}{2}+
\\ \nn
&&\hspace*{4cm}+\int \dmza
\log\left(2\cosh z_0 \sqrt{\lambda_0}\right)
\EEA
and the final form of the rescaled free energy, as defined in (\ref{energy}),
is given by
\BEQ
f(\mu,\alpha)=\hspace*{-0.8mm}-\hspace*{-0.8mm}\frac{\mu}{16}\hspace*{-0.8mm}
\left[1\hspace*{-0.8mm}-\hspace*{-0.8mm}\frac{(1+q_0)^2(2-q_0)}{4}\right]
\hspace*{-0.8mm}+\hspace*{-0.8mm}\frac{1}{\mu}\int\hspace*{-0.8mm} \dmza\hspace*{-0.8mm}
\log\hspace*{-0.8mm}\left(2\cosh z_0 \sqrt{\lambda_0}\right) \ ,
\EEQ
where the parameters $q_{0}$ and $\lambda_{0}$  satisfy the equations:
\BEQ
\lambda_0=\frac{3}{64}\mu^2(1+q_0)^2\ , \ \ \ \ 
q_0=\int\dmza \left(\tanh(z\sqrt{\lambda_0})\right)^2 \ ,
\label{eq:q0}
\EEQ
\noindent and we have used the following compact measure for the Gaussian measure:
\BEQ
dp(z) \equiv\frac{e^{-z^{2}/2}}{\sqrt{2\pi}} dz \ .
\EEQ

The solution of the equation (\ref{eq:q0}) can be found by iterations.
 The parameter $q_{0}$ is 
an analytic function of $\tau$ (fig.
3, full curve). 
No transition is present. 
The zero temperature value of the
rescaled energy  (\ref{energy}) is given by $e_{0}= \sqrt{\frac{3}{8 \pi}} 
\approx .345494$.

The corresponding value of $\alpha$ where the energy density 
$E(\mu,\alpha)$ goes to 
zero, i. e. the ratio of the number of variables and the number of clauses 
up to which  all clauses are satisfied, turns out to be, following 
(\ref{STIMA}),
 $\alpha_c=24/\pi=7.6394373$ in this asymptotic approximation.

The value of the entropy corresponding to this solution is shown in figure
1. 
It becomes negative at low temperature  signaling an
 inconsistency of the 
approach based on replica symmetry. 
In order to obtain reasonable
 results
in the low temperature region we must break the replica symmetry as
we will show in the next section.

\begin{figure}[!htb]
\begin{center}
\includegraphics[width=0.48\textwidth,height = 0.32\textheight]{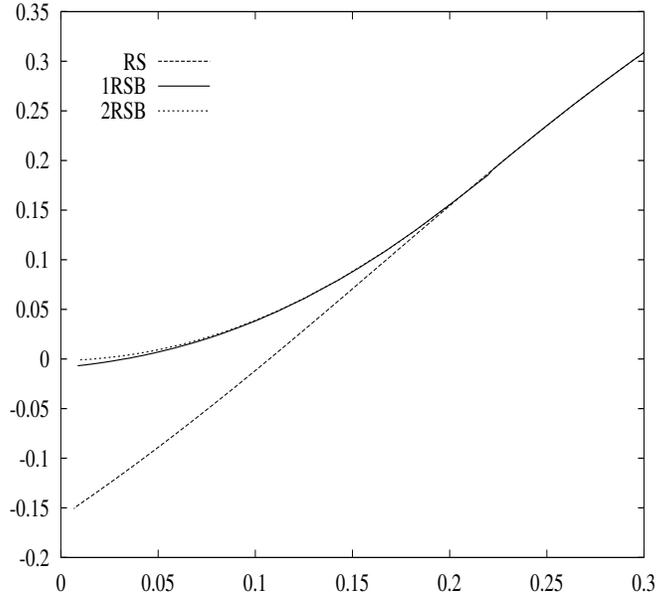} 
 \protect\caption{  Entropy as function of the reduced temperature 
$\tau$ in the replica symmetric case  and in the broken symmetry
 case with one and two steps 
breaking.} 
\label{ENTROPY}
\end{center}
\end{figure}

\begin{figure}[!htb]
\begin{center}
\includegraphics[width=0.48\textwidth,height = 0.32\textheight]{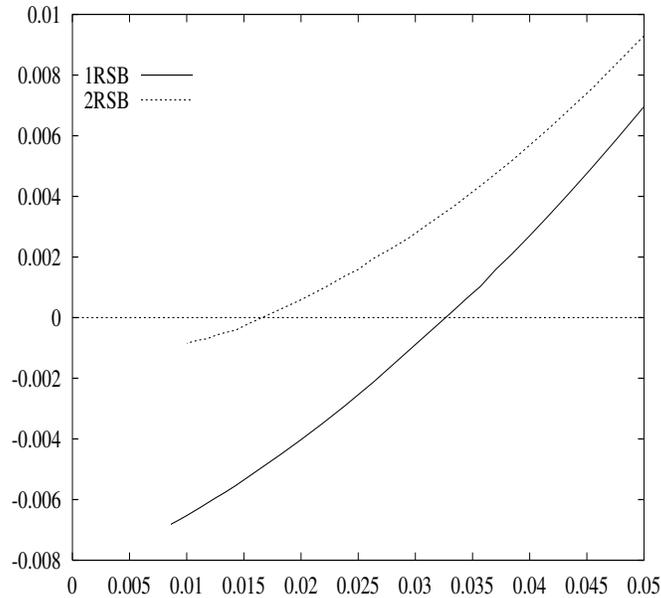}
 \protect\caption{Detail of the cases with broken replica symmetry,
  shown in figure \ref{ENTROPY}, at low temperature.
}
\label{ENTROPY_det}
\end{center}
\end{figure}

\section{The Replica Symmetry Breaking}
\subsection{One step Replica Symmetry Breaking}
If replica symmetry is broken, very often reasonable results are obtained 
in the framework of the 
one step replica symmetry breaking, where it is assumed that the elements 
of the matrix $Q$ take 
only two values (for the physical interpretation of one step replica symmetry
 breaking see reference.  
\cite{mpv}).

In the one step case one divides the indices $a$ in $n/m$ groups,
 each group having $m$ components.  
We set $Q_{a,b}$ equal to $q_{1}$ if $a$ and $b$ belong to the same
 group, otherwise we set 
$Q_{a,b}$ equal to $q_{0}$.  Similar relations are used for the matrix 
$\Lambda$.  The free energy 
is now a function of three independent parameters: $m$, $q_0$ and $q_1$. 
 The limit $n\to 0$ is 
obtained in this case by doing an analytic continuation also in $m$, that
 in such a process will not 
be integer anymore.  In the $0$ replicae limit $m$ acquires non-integer 
values between $0$ and 
$1$.

After some simple computation we get
\BEA
&&A(q_0,q_1;\lambda_0,\lambda_1,m)=
\\
\nn
&&\hspace{0.7 cm}=-\frac{\mu\sqrt{\alpha}}{8}+\frac{\mu^2}{16}-\frac{\mu^2}{128}\left[
m(1+q_0)^3+(1-m)(1+q_1)^3\right]+
\\ \nn
&&\hspace*{0.7 cm}-\frac{\lambda_1}{2}+\mez [m\lambda_0q_0+(1-m)\lambda_1q_1]+
\\ \nn
&&\hspace*{0.7 cm}+\frac{1}{m}\hspace*{-0.5mm}\int \hspace*{-0.5mm}\dmza\hspace*{-0.5mm}
\log \hspace*{-0.5mm}
\left(\hspace*{-0.5mm}\int\hspace*{-0.5mm}\dmzb \hspace*{-0.5mm}
\left(\hspace*{-0.5mm}
2\cosh \hspace*{-0.5mm}\left(\hspace*{-0.5mm}
	z_0 \sqrt{\lambda_0}\hspace*{-0.5mm}+\hspace*{-0.5mm}
	z_1 \sqrt{\lambda_1-\lambda_0}
             \hspace*{-0.5mm}   \right)\hspace*{-0.5mm}
	 \right)^m\hspace*{-0.5mm} \right) \ ,
\EEA
\BEA
&&f(\mu)=
\\
\nn
&&\hspace*{0.7 cm}-\frac{\mu}{16}\left[
1-\frac{m}{8}\left((1+q_0)^2(1-2q_0)-(1+q_1)^2(1-2q_1)\right)+\right.
\\
\nn
&&\hspace*{0.7 cm}\left.-\frac{(1+q_1)^2(2-q_1)}{4}\right]+
\\ 
\nn
&&\hspace*{0.7 cm} -\frac{1}{m\mu}\hspace*{-0.5mm}\int\hspace*{-0.5mm} \dmza\hspace*{-0.5mm}
\log\hspace*{-0.5mm}\int\hspace*{-0.5mm}\dmzb \hspace*{-0.5mm}
\left(\hspace*{-0.5mm}2\cosh \hspace*{-0.5mm}\left(\hspace*{-0.5mm}
	z_0 \sqrt\lambda_0\hspace*{-0.5mm}+\hspace*{-0.5mm}
	 z_1 \sqrt{\lambda_1-
\lambda_0}\hspace*{-0.5mm}\right)\hspace*{-0.5mm}\right)^m \ ,
\EEA
where the following equations are satisfied:

\BEQ
\lambda_i=\frac{3}{64}\mu^2(1+q_i)^2, \hspace*{1cm} i=0,1
\EEQ
\BEA
q_0&=&\int\dmza \left( 
\frac{1}{\int\dmzb
\left(\cosh \left(z_0 \sqrt\lambda_0+ z_1 \sqrt{\lambda_1-
\lambda_0}\right)\right)^m}\right.\times
\\ \nn
&\times&\hspace*{-4mm}\left.\hspace*{-0.8mm}\int\hspace*{-0.8mm}\dmzb\hspace*{-0.8mm}
 \tanh\hspace*{-0.8mm}\left(
z_0\sqrt{\lambda_0}\hspace*{-0.8mm}+\hspace*{-0.8mm}
z_1 \sqrt{\lambda_1\hspace*{-0.8mm}-\hspace*{-0.8mm}\lambda_0}\right)
\hspace*{-0.8mm}\left(\hspace*{-0.8mm}\cosh \hspace*{-0.8mm}
\left(z_0 \sqrt\lambda_0\hspace*{-0.8mm}+\hspace*{-0.8mm}
 z_1 \sqrt{\lambda_1\hspace*{-0.8mm}-\hspace*{-0.8mm}
\lambda_0}\hspace*{-0.8mm}\right)\hspace*{-0.8mm}\right)^m\hspace*{-0.8mm}
\right)^2
\\ 
q_1&=&\int\dmza \frac{1}
{\int\dmzb
\left(\cosh \left(z_0 \sqrt\lambda_0+ z_1 \sqrt{\lambda_1-
\lambda_0}\right)\right)^m}\times
\\ \nn
&\times&
\hspace*{-4mm}\int\hspace*{-0.8mm}\dmzb \hspace*{-0.8mm}\left(\hspace*{-0.8mm}
\tanh\hspace*{-0.8mm}\left(z_0\sqrt{\lambda_0}\hspace*{-0.8mm}+\hspace*{-0.8mm}
z_1 \sqrt{\lambda_1\hspace*{-0.8mm}-\hspace*{-0.8mm}\lambda_0}
\hspace*{-0.8mm}\right)\hspace*{-0.8mm}\right)^2\hspace*{-0.8mm}
\left(\hspace*{-0.8mm}\cosh \hspace*{-0.8mm}
\left(z_0 \sqrt\lambda_0\hspace*{-0.8mm}+\hspace*{-0.8mm}
 z_1 \sqrt{\lambda_1\hspace*{-0.8mm}-\hspace*{-0.8mm}
\lambda_0}\hspace*{-0.8mm}\right)\hspace*{-0.8mm}\right)^m
\EEA
and the parameter $m$ is chosen in such a way to minimize
 the resulting free energy.

One finds that for $\mu>\mu_{c}=4.55 $ the previous equations have
 a non trivial 
solution (e.g. $m\ne0,q_1\neq q_0$). The corresponding values 
of $m$, $q_{0}$ and $q_{1}$ 
are shown 
in figures 3 and 4.
 It is evident that the value of $m$ 
is different from the 
one at  critical temperature and that the difference $q_{1}-q_{0}$ 
vanishes at $\mu_c$. 
This behavior is similar to the one that is realized in the 
Sherrington-Kirkpatrick (SK) model in non-zero magnetic field.

\begin{figure}[!htb] 
\begin{center}
\includegraphics[width=0.48\textwidth,height = 0.32\textheight]{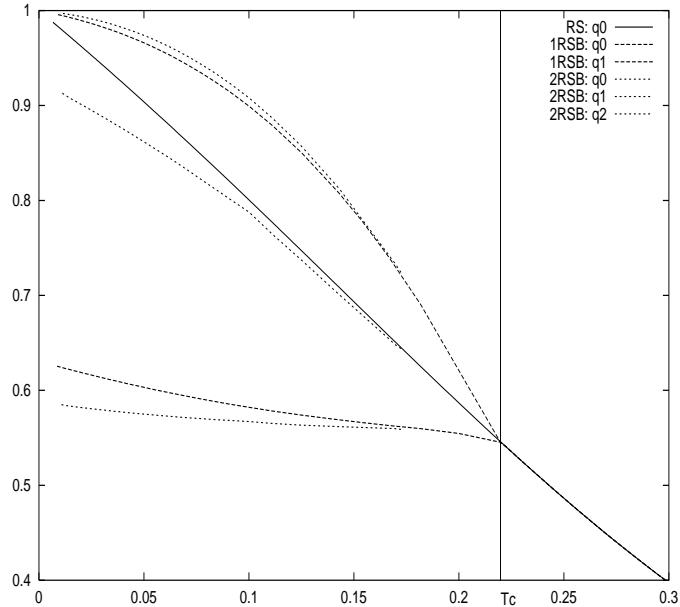}
 \protect\caption{Values of $q_{0}$ as function of 
the temperature $\tau$ in the replica symmetric case (full curve), 
 of $q_{0}$ and $q_{1}$ in the 
broken symmetry case with one step breaking (dashed curves)
and those of $q_0$, $q_1$ and $q_2$ in the
 two step replica symmetry breaking case (dotted curves).} \label{Q}
\end{center}
\end{figure}
\begin{figure}[!htb]
\begin{center}
\includegraphics[width=0.48\textwidth,height = 0.32\textheight]{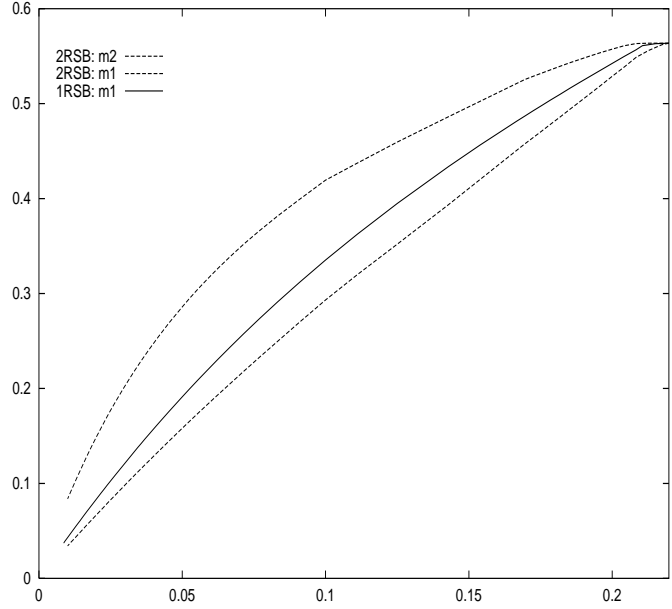}
 \protect\caption{Parameters $m$'s as function of $\tau$
for one step (full curve)
and two steps (dashed curves)  replica symmetry breaking.}
 \label{M}
\end{center}
\end{figure}

In figures 1 and 2
 we plot the entropy as function of the temperature. 
Also in this case 
the entropy becomes negative at sufficiently small temperature, 
but this happens in a 
rather smaller region above $\tau=0$. 

In the SK model for spin glasses, where this disaster happens at one step
 level, (we recall that the 
entropy cannot be negative), the correct value of  
the entropy is proportional to $\tau^{2}$.  If a similar behaviour is
 present in this model the free 
energy should be given by $A+B\tau^{3}$ at small, but not too small $\tau$. 
 We show in figures 5 
and 6 the free energy as function of $ \tau^{3}$.  We see that in a wide
 range of $\tau^{3}$ a 
linear behaviour is present supporting a quadratic dependence of the entropy 
on the the temperature.  
The extrapolated value of the zero temperature  rescaled  free energy 
obtained using this 
method ($f_{ext}^{(1)}=A=-0.333412$) is slightly larger than the actual value 
at zero temperature 
($f_{\mbox{\tiny{$1RSB$}}}(T=0)=-0.333740$), however this first value should 
be more reliable, 
because it is known that the errors in the free energy, if one makes the
 approximation of 
considering only a finite number of RSB steps, are negligible at higher 
temperature, but they become
larger at low temperature. $f_{ext}^{(1)}$
  gives an estimate (\ref{STIMA}) equal to 
$\alpha_c\simeq 7.114468$

\begin{figure}[!htb]
\begin{center}
 \includegraphics[width=0.48\textwidth,height = 0.32\textheight]{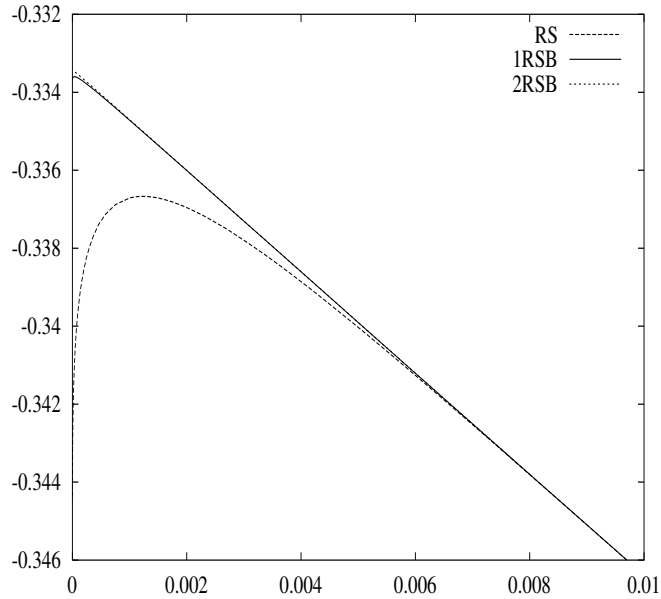}
 \protect\caption{Rescaled free energy  
versus 
the cube of the reduced temperature $\tau^{3}$  in the replica symmetric case
and in the
 broken symmetry cases with one and two steps breaking.}
\label{FREE}
\end{center}
\end{figure}
\begin{figure}[!htb]
\begin{center}
\includegraphics[width=0.48\textwidth,height = 0.32\textheight]{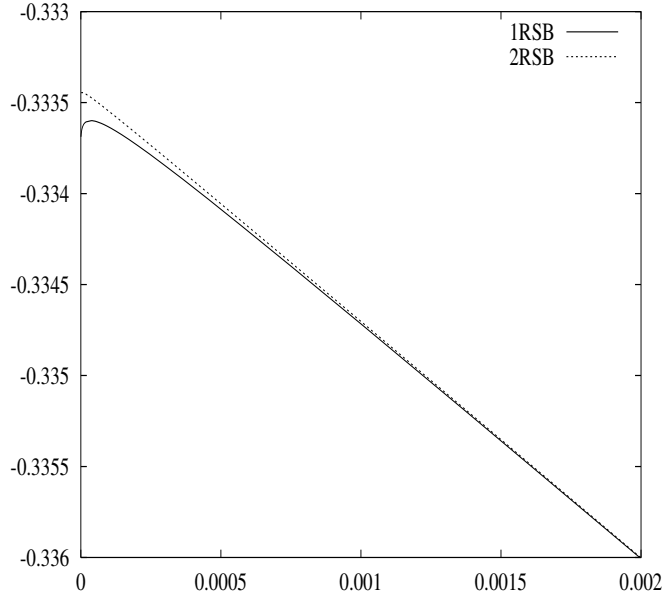}
 \protect\caption{Detail of figure \ref{FREE} showing
the one and two step symmetry breaking cases at low $\tau$.
}
\label{FREE_det}
\end{center}
\end{figure}

\subsection{Two steps Replica Symmetry Breaking}

In we perform another step in breaking the replica symmetry and we let
the elements of
$Q$ and $\Lambda$ take three different values ($q_0$, $q_1$ and $q_2$
and $\lambda_0$, $\lambda_1$ and $\lambda_2$ respectively)
the free energy will be a function of the five independent variables:
$q_0$, $q_1$, $q_2$, $m_1$, $m_2$.
In this case we have:
\BEA
&&\hspace*{-0.7cm} A(q_0,q_1,q_2;\lambda_0(q_0),\lambda_1(q_1),\lambda_2(q_2))= -\frac{\mu\sqrt{\alpha}}{8}+\frac{\mu^2}{16}-
\\
\nonumber
&&-\frac{\mu^2}{128}\left[
m_1(1+q_0)^3+(m_2-m_1)(1+q_1)^3+(1-m_2)(1+q_2)^3\right]+
\\ \nonumber
&&-\frac{\lambda_2}{2}+\mez [m_1\lambda_0q_0+(m_2-m_1)\lambda_1q_1+
(1-m_2)\lambda_2q_2]+
\\ \nonumber
&&+\frac{1}{m_1}\int \dmza \times
\\
\nn
&&\hspace*{- 0.7cm}\times\hspace*{-1mm}
\log \hspace*{-1mm}
\left(\hspace*{-1mm}\int\hspace*{-1mm}\dmzb \hspace*{-1mm}
\left[\hspace*{-1mm}\int\hspace*{-1mm}\dmzc\hspace*{-1mm}
\left(\hspace*{-1mm}
2\hspace*{-0.6mm}\cosh \hspace*{-1mm}\left(\hspace*{-1mm}
z_0\hspace*{-0.4mm}
 \sqrt{\hspace*{-0.4mm}\lambda_0}\hspace*{-1mm}+\hspace*{-1mm}
z_1\hspace*{-0.4mm} \sqrt{\hspace*{-0.4mm}
\lambda_1\hspace*{-1mm}-\hspace*{-1mm}\lambda_0}
\hspace*{-1mm}+\hspace*{-1mm}
z_2 \hspace*{-0.4mm}\sqrt{\hspace*{-0.4mm}
\lambda_2\hspace*{-1mm}-\hspace*{-1mm}\lambda_1}
	\hspace*{-1mm}\right)
	\hspace*{-1mm}\right)^{m_2}
	\hspace*{-1mm}\right]\hspace*{-1mm}^{\frac{m_1}{m_2}}
	\hspace*{-1mm}\right) \ ,
\EEA
\BEA
&&\hspace*{- 0.7cm} f(\mu)=-\frac{\mu}{16}\left[
1-\frac{1}{8}\left( m_1\left((1+q_0)^2(1-2q_0)-(1+q_1)^2(1-2q_1)\right)+
\right.\right.
\\
\nonumber
&& \left.\left.+\hspace*{- 0.4mm}
m_2\left(\hspace*{- 0.4mm}(1\hspace*{- 0.4mm}+\hspace*{- 0.4mm}q_1)^2
(1\hspace*{- 0.4mm}-\hspace*{- 0.4mm}2q_1)
\hspace*{- 0.4mm}-\hspace*{- 0.4mm}
(1\hspace*{- 0.4mm}+\hspace*{- 0.4mm}q_2)^2(1\hspace*{- 0.4mm}-
\hspace*{- 0.4mm}2q_2)
\hspace*{- 0.4mm}\right)\hspace*{- 0.4mm}\right)\hspace*{- 0.4mm}
-\hspace*{- 0.4mm}
\frac{(1\hspace*{- 0.4mm}+\hspace*{- 0.4mm}q_2)^2(2\hspace*{- 0.4mm}-\hspace*{- 0.4mm}q_2)}{4}\right]+
\\ 
\nonumber
&& -\frac{1}{m_1\mu}\int \dmza \times
\\ \nn
&&\hspace*{- 0.7cm}\times\hspace*{-1mm}
\log \hspace*{-1mm}
\left(\hspace*{-1mm}\int\hspace*{-1mm}\dmzb \hspace*{-1mm}
\left[\hspace*{-1mm}\int\hspace*{-1mm}\dmzc\hspace*{-1mm}
\left(\hspace*{-1mm}
2\hspace*{-0.6mm}\cosh \hspace*{-1mm}\left(\hspace*{-1mm}
z_0\hspace*{-0.4mm}
 \sqrt{\hspace*{-0.4mm}\lambda_0}\hspace*{-1mm}+\hspace*{-1mm}
z_1\hspace*{-0.4mm} \sqrt{\hspace*{-0.4mm}
\lambda_1\hspace*{-1mm}-\hspace*{-1mm}\lambda_0}
\hspace*{-1mm}+\hspace*{-1mm}
z_2 \hspace*{-0.4mm}\sqrt{\hspace*{-0.4mm}
\lambda_2\hspace*{-1mm}-\hspace*{-1mm}\lambda_1}
	\hspace*{-1mm}\right)
	\hspace*{-1mm}\right)^{m_2}
	\hspace*{-1mm}\right]\hspace*{-1mm}^{\frac{m_1}{m_2}}
	\hspace*{-1mm}\right) \ ,
\EEA
\noindent where the self consistency equations are: 
\BEQ
 \lambda_i=\frac{3}{64}\mu^2(1+q_i)^2,\hspace*{ 1cm}
i=0,1,2
\EEQ

{\footnotesize{
\BEA
&&\hspace*{-1cm} \nn
q_0\hspace*{1mm}
=\hspace*{-1.2mm}\int\hspace*{-1.2mm}\dmza\hspace*{-1.2mm}
\left(\hspace*{-0.7mm}
\frac{1}{\int\hspace*{-0.7mm}\dmzb\hspace*{-0.7mm}
\left[\hspace*{-0.7mm}\int\hspace*{-0.7mm}\dmzc\hspace*{-0.7mm}
\left(\hspace*{-0.7mm}\cosh \hspace*{-0.7mm}\left(\hspace*{-0.7mm}
	z_0\hspace*{-0.7mm} \sqrt{\hspace*{-0.7mm}}\lambda_0
	\hspace*{-0.7mm}+\hspace*{-0.7mm}
	 z_1\hspace*{-0.7mm} \sqrt{\hspace*{-0.7mm}
	\lambda_1\hspace*{-0.7mm}-\hspace*{-0.7mm}\lambda_0}
	\hspace*{-0.7mm}+\hspace*{-0.7mm}
	 z_2 \hspace*{-0.7mm}\sqrt{\hspace*{-0.7mm}
	\lambda_2\hspace*{-0.7mm}-\hspace*{-0.7mm}\lambda_1}
	\hspace*{-0.7mm}\right)\hspace*{-0.7mm}\right)\hspace*{-0.7mm}^{m_2}
	 \hspace*{-0.7mm}\right]
\hspace*{-0.7mm}^{\frac{m_1}{m_2}}}\hspace*{-0.7mm}
	\times\right.
\\
\nn
&&\hspace*{0.1 cm}
\times \left.
\hspace*{-0.7mm}\int\hspace*{-0.7mm}\dmzb \hspace*{-0.7mm}
\left[\hspace*{-0.7mm}\int\hspace*{-0.7mm}\dmzc\hspace*{-0.7mm}
 \left(\hspace*{-0.7mm}\cosh \hspace*{-0.7mm}
\left(\hspace*{-0.7mm}
	z_0\hspace*{-0.7mm} \sqrt{\hspace*{-0.7mm}
	\lambda_0}\hspace*{-0.7mm}+ \hspace*{-0.7mm}
	z_1\hspace*{-0.7mm} \sqrt{\hspace*{-0.7mm}
	\lambda_1\hspace*{-0.7mm}-\hspace*{-0.7mm}\lambda_0}
	\hspace*{-0.7mm}+\hspace*{-0.7mm} 
	z_2 \hspace*{-0.7mm}\sqrt{\hspace*{-0.7mm}
	\lambda_2\hspace*{-0.7mm}-\hspace*{-0.7mm}\lambda_1}
	\hspace*{-0.7mm}\right)\hspace*{-0.7mm}\right)\hspace*{-0.7mm}^{m_2} 
\hspace*{-0.7mm}\right]\hspace*{-0.7mm}^{\frac{m_1}{m_2}-1}
\hspace*{-0.7mm}\times \right.
\\ 
 &&\hspace{1.7cm}\left.
 \times\hspace{-0.7mm}\int\hspace{-0.7mm}\dmzc\hspace{-0.7mm}
\tanh\hspace{-0.7mm}\left(\hspace{-0.7mm}
	z_0 \hspace{-0.7mm}\sqrt{\hspace{-0.7mm}\lambda_0}
	\hspace{-0.7mm}+\hspace{-0.7mm}
	 z_1\hspace{-0.7mm}\sqrt{\hspace{-0.7mm}\lambda_1\hspace{-0.7mm}
	-\hspace{-0.7mm}\lambda_0}\hspace{-0.7mm}+\hspace{-0.7mm}
	 z_2\hspace{-0.7mm} \sqrt{\hspace{-0.7mm}\lambda_2\hspace{-0.7mm}
	-\hspace{-0.7mm}\lambda_1}\hspace{-0.7mm}\right) \hspace{-0.7mm}
	\right.\times
\\
\nn
&&\hspace{3.1cm}\times\left.
\left(\hspace{-0.7mm}\cosh \hspace{-0.7mm}
\left(\hspace{-0.7mm}
	z_0\hspace{-0.7mm}\sqrt{\hspace{-0.7mm}\lambda_0}
	\hspace{-0.7mm}+\hspace{-0.7mm}
	 z_1\hspace{-0.7mm}\sqrt{\hspace{-0.7mm}\lambda_1
	\hspace{-1 mm}-\hspace{-0.7mm}\lambda_0}
	\hspace{-0.7mm}+ \hspace{-0.7mm}
	z_2\hspace{-0.7mm} \sqrt{\hspace{-0.7mm}\lambda_2\hspace{-0.7mm}
	-\hspace{-0.7mm}\lambda_1}\hspace{-0.7mm}\right)
	\hspace{-0.7mm}\right)^{m_2}
	\hspace{-0.7mm} 
\right)^2
\EEA
}}
{\footnotesize{
\BEA
&&\hspace*{-1cm} \nn
q_1\hspace*{-1.2mm} = \hspace*{-1.2mm}\int\hspace*{-1.2mm}
\dmza \hspace*{-1.2mm}\left(\hspace*{-0.7mm}
\frac{1}{\int\hspace*{-0.7mm}\dmzb\hspace*{-0.7mm}
\left[\hspace*{-0.7mm}\int\hspace*{-0.7mm}\dmzc\hspace*{-0.7mm}
\left(\hspace*{-0.7mm}\cosh\hspace*{-0.7mm} \left(\hspace*{-0.7mm}
	z_0\hspace*{-0.7mm} \sqrt{\hspace*{-0.7mm}\lambda_0}
	\hspace*{-0.7mm}+\hspace*{-0.7mm}
	 z_1\hspace*{-0.7mm} \sqrt{\hspace*{-0.7mm}
	\lambda_1\hspace*{-0.7mm}-\hspace*{-0.7mm}\lambda_0}
	\hspace*{-0.7mm}+\hspace*{-0.7mm}
	 z_2 \hspace*{-0.7mm}\sqrt{\hspace*{-0.7mm}\lambda_2
	\hspace*{-0.7mm}-\hspace*{-0.7mm}\lambda_1}
	\hspace*{-0.7mm}\right)\hspace*{-0.7mm}\right)\hspace*{-0.7mm}^{m_2}
\hspace*{-0.7mm} \right]\hspace*{-0.7mm}^{\frac{m_1}{m_2}}}
\hspace*{-0.7mm}\times\right.
\\
\nn
&&\hspace*{1mm}\left.\int\hspace*{-0.7mm}\dmzb 
\hspace*{-0.7mm}\left[
\hspace*{-0.7mm}\int\hspace*{-0.7mm}\dmzc\hspace*{-0.7mm}
 \left(\hspace*{-0.7mm}\cosh \hspace*{-0.7mm}
\left(\hspace*{-0.7mm}
	z_0\hspace*{-0.7mm} \sqrt{\hspace*{-0.7mm}\lambda_0}
	\hspace*{-0.7mm}+\hspace*{-0.7mm}
	 z_1\hspace*{-0.7mm} \sqrt{\hspace*{-0.7mm}
	\lambda_1\hspace*{-0.7mm}-\hspace*{-0.7mm}\lambda_0}
	\hspace*{-0.7mm}+\hspace*{-0.7mm}
	 z_2 \hspace*{-0.7mm}\sqrt{\hspace*{-0.7mm}\lambda_2
	\hspace*{-0.7mm}-\hspace*{-0.7mm}\lambda_1}
	\hspace*{-0.7mm}\right)\hspace*{-0.7mm}\right)\hspace{-0.7mm}^{m_2}
	\hspace{-0.7mm} \right]^{\frac{m_1}{m_2}-2} \right.
\hspace*{-0.7mm}\hspace*{-0.7mm}\times
\\  
&&\hspace*{1.3cm} \left.\times\hspace{-0.7mm} 
\left[\int\hspace{-0.7mm}\dmzc
 \hspace{-0.7mm}\left(\cosh \hspace{-0.7mm}\left(
	z_0\hspace{-0.7mm} \sqrt{\hspace{-0.7mm}\lambda_0}
	\hspace{-0.7mm} +\hspace{-1 mm} 
	z_1\hspace{-0.7mm} \sqrt{\hspace{-0.7mm}\lambda_1\hspace{-0.7mm}-
	\hspace{-0.7mm}\lambda_0}
	\hspace{-0.7mm}+ \hspace{-0.7mm} 
	z_2\hspace{-0.7mm} \sqrt{\hspace{-0.7mm}\lambda_2\hspace{-0.7mm}-
	\hspace{-0.7mm}\lambda_1}
	\hspace{-0.7mm}\right)\hspace{-0.7mm}\right)\hspace{-0.7mm}^{m_2}
\hspace{-0.7mm}\times\right.\right.
\\\nonumber
&&\hspace*{3.1cm}\times \left.\left.\hspace{-0.7mm}\tanh\hspace{-0.7mm}\left(
	z_0\hspace{-0.7mm}\sqrt{\hspace{-0.7mm}\lambda_0}
	\hspace{-0.7mm} + \hspace{-0.7mm} 
	z_1\hspace{-0.7mm}\sqrt{\hspace{-0.7mm}\lambda_1\hspace{-0.7mm}-
	\hspace{-0.7mm} \lambda_0}
	\hspace{-0.7mm} +\hspace{-1 mm}  
	z_2\hspace{-0.7mm}  \sqrt{\hspace{-0.7mm}\lambda_2\hspace{-0.7mm}-
	\hspace{-0.7mm} \lambda_1}
	\hspace{-0.7mm}\right)
\hspace{-0.7mm}\right]\hspace{-0.7mm}^2\hspace{-0.7mm}
\right)
\EEA
}}
{\footnotesize{
\BEA
&&\hspace*{-1cm} \nn 
q_2\hspace*{-0.7mm}=\hspace*{-0.7mm}\int\hspace*{-0.7mm}\dmza\hspace*{-0.7mm} \left(\hspace*{-0.7mm}
\frac{1}
{\int\hspace*{-0.7mm}\dmzb\hspace*{-0.7mm}
\left[\hspace*{-0.7mm}\int\hspace*{-0.7mm}\dmzc\hspace*{-0.7mm}
\left(\hspace*{-0.7mm}\cosh\hspace*{-0.7mm} \left(\hspace*{-0.7mm}
	z_0\hspace*{-0.7mm} \sqrt{\hspace*{-0.7mm}\lambda_0}
	\hspace*{-0.7mm}+\hspace*{-0.7mm}
	 z_1 \hspace*{-0.7mm}\sqrt{\hspace*{-0.7mm}\lambda_1\hspace*{-0.7mm}-
	\hspace*{-0.7mm}\lambda_0}
	\hspace*{-0.7mm}+\hspace*{-0.7mm}
	 z_2 \hspace*{-0.7mm}\sqrt{\hspace*{-0.7mm}\lambda_2\hspace*{-0.7mm}-
	\hspace*{-0.7mm}\lambda_1}
	\hspace*{-0.7mm}\right)\hspace*{-0.7mm}\right)\hspace*{-0.7mm}^{m_2}
	\hspace*{-0.7mm} \right]\hspace*{-0.7mm}^{\frac{m_1}{m_2}}}\hspace*{-0.7mm}\times
\right.
\\
\nn
&&\left.\times\hspace*{-0.7mm}\int\hspace*{-0.7mm}\dmzb\hspace*{-0.7mm}
 \left[\hspace*{-0.7mm}\int\hspace*{-0.7mm}\dmzc \hspace*{-0.7mm}
\left(\hspace*{-0.7mm}\cosh \hspace*{-0.7mm}\left(\hspace*{-0.7mm}
	z_0 \hspace*{-0.7mm}\sqrt{\hspace*{-0.7mm}\lambda_0}
	\hspace*{-0.7mm}+\hspace*{-0.7mm}
	 z_2\hspace*{-0.7mm} \sqrt{\hspace*{-0.7mm}\lambda_1\hspace*{-0.7mm}-
	\hspace*{-0.7mm}\lambda_0}
	\hspace*{-0.7mm}+\hspace*{-0.7mm}
	 z_1 \hspace*{-0.7mm}\sqrt{\hspace*{-0.7mm}\lambda_2\hspace*{-0.7mm}-
	\hspace*{-0.7mm}\lambda_1}
	\hspace*{-0.7mm}\right)\hspace*{-0.7mm}\right)\hspace*{-0.7mm}^{m_2}
	 \hspace*{-0.7mm} \right]^{\frac{m_1}{m_2}-1}\right.
\hspace*{-0.7mm}\hspace*{-0.7mm}\times
\\ 
&&\hspace*{1.5cm} \left.\times\hspace{-0.7mm}
\int\hspace{-0.7mm}\dmzc \hspace{-0.7mm}\left(\cosh\hspace{-0.7mm}
 \left(
	z_0\hspace{-0.7mm} \sqrt{\hspace*{-0.7mm}\lambda_0}
	\hspace{-0.7mm}+\hspace{-0.7mm} 
	z_1 \hspace{-1 mm}\sqrt{\hspace*{-0.7mm}\lambda_1\hspace{-0.7mm}-
	\hspace{-0.7mm}\lambda_0}
	\hspace{-0.7mm}+\hspace{-0.7mm} 
	z_2\hspace{-0.7mm} \sqrt{\hspace*{-0.7mm}\lambda_2\hspace{-0.7mm}-
	\hspace{-1 mm}\lambda_1}
	\hspace*{-0.7mm}\right)\hspace*{-0.7mm}\right)\hspace*{-0.7mm}^{m_2} 
\hspace{-0.7mm}\times\right.
\\
\nn
&&\hspace*{3 cm}\times\hspace*{-0.7mm\left.\left(\hspace{-0.7mm}\tanh \hspace{-0.7mm}\left(\hspace{-0.7mm}
	z_0\hspace{-0.7mm} \sqrt{\hspace*{-0.7mm}\lambda_0}
	\hspace{-0.7mm}+\hspace{-0.7mm} 
	z_1 \hspace{-1 mm}\sqrt{\hspace*{-0.7mm}\lambda_1\hspace{-0.7mm}-
	\hspace{-0.7mm}\lambda_0}
	\hspace{-0.7mm}+\hspace{-0.7mm} 
	z_2\hspace{-0.7mm} \sqrt{\hspace*{-0.7mm}\lambda_2\hspace{-0.7mm}-
	\hspace{-1 mm}\lambda_1}
	\hspace{-0.7mm}\right)\hspace{-0.7mm}\right)\hspace{-0.7mm}^2
	\hspace{-1 mm}
\right)
\EEA
}}}

From figures 3 and 4 we see that the transition between the replica 
symmetric structure and the 
broken one at $\tau_c=1/\mu_c=0.21978$ is confirmed.  In the two step
 computation the entropy (fig.  
2) still becomes negative but at a lower temperature than in the one step 
case and the zero 
temperature value is less negative than before.  Even the $\tau^2$ 
behaviour for small $\tau$, or 
the equivalent $A+B\tau^3$ law for the free energy, is satisfied up 
to a smaller temperature.  The 
extrapolated value of the zero temperature free energy is now 
$f_{ext}^{(2)}=A=-0.333401$, where the 
value given by the actual 2RSB free energy is 
$f_{\mbox{\tiny{$2RSB$}}}=-0.333450$.  Their 
difference is of an order of magnitude smaller than in the one step 
replica symmetry breaking case 
(we recall that in that case $f_{ext}^{(1)}=-0.333412$ and 
$f_{\mbox{\tiny{$1RSB$}}}(T=0)=-0.333740$). 

 The pathologies here exhibited
clearly show that the one step and the two steps solutions are still 
 not exact, and it is natural to suppose that they will disappear when 
we break the replica 
symmetry in a continuous way \cite{mpv}.

Using 
 equation (\ref{STIMA}) once again,
we find
$\alpha_{c}\approx 7.1139985$.
This value
  is not too far from the estimated result (i.e. 
4.2 \cite{KISE}) if we consider how crude is  our approximation.


\section{Conclusions}

We have seen that the transition from the replica symmetric case to
 the replica broken 
case is a smooth transition,
 which is quite different from the quasi first order transition 
of the $p$-spin model \cite{KWT}.  
This difference is likely due to the fact that the self overlap 
$q_{0}$ is different from zero  in the high temperature phase. 

The one step approximation is not exact at low temperature and
 it is likely not to be exact 
in the whole replica symmetry broken phase. The two step replica symmetry 
breaking computation gives clear hints
that the replica symmetry must be broken in a continuous way. In any case 
both approximations 
apparently give excellent approximations for the free energy at low 
temperature (the error on the value of the 
zero temperature free energy is $O(10^{-4})$ at one step and $O(10^{-5})$ 
at two steps RSB level 
respectively). 

The model has a behaviour that is  very similar 
to the one  of the Sherrington-Kirkpatrick
model in presence of a magnetic field.
It is natural to conjecture that these properties hold in a
quite large interval  
of values of $\alpha$,  for $\alpha \geq \alpha_c$, even far from
the full connectivity limit $\alpha \to \infty$.
It would be very interesting to check these predictions 
using numerical simulations and/or
analytic tools.

We wish to thank Enzo Marinari and Riccardo Zecchina 
for many useful discussions on the 
subject of this paper.



\end{document}